# Three-dimensional thermographic imaging using a virtual wave concept

Peter Burgholzer,[1, a)] Michael Thor,[2] Jürgen Gruber,[2] and Günther Mayr[2]

[1] *Christian Doppler Laboratory for Photoacoustic Imaging and Laser Ultrasonics, Research Center for Non Destructive Testing (RECENDT), 4040 Linz, Austria*

[2] *School of Engineering, University of Applied Sciences Upper Austria, 4600 Wels, Austria*

In this work, it is shown that image reconstruction methods from ultrasonic imaging can be employed for thermographic signals. Before using these imaging methods, a virtual signal is calculated by applying a local transformation to the temperature evolution measured on a sample surface. The introduced transformation describes all the irreversibility of the heat diffusion process and can be used for every sample shape. To date, one-dimensional methods have been primarily used in thermographic imaging. The proposed two-stage algorithm enables reconstruction in two and three dimensions. The feasibility of this approach is demonstrated through simulations and experiments. For the latter, small steel beads embedded in an epoxy resin are imaged. The resolution limit is found to be proportional to the depth of the structures and to be inversely proportional to the logarithm of the signal-to-noise ratio. Limited-view artefacts can arise if the measurement is performed on a single planar detection surface. These artifacts can be reduced by measuring the thermographic signals from multiple planes, which is demonstrated by numerical simulations and by experiments performed on an epoxy cube.

## I. INTRODUCTION

Active thermography is used for detection, imaging, and sizing of structural imperfections in a non-destructive way[1]. Applications range from diagnosis of sub-surface defects and material characterization up to on-line and in-line inspection[2]. Active thermography involves heating or cooling of a sample to create a temperature gradient and observation of the heat diffusion process by measuring the infrared thermal signature of the sample using an infrared camera. To heat up the sample various methods are used, including a flash-lamp or laser-heating[3,4,5,6], ultrasonic waves[7], or inductive heating[8]. Various temporal excitation methods, such as pulses in pulsed thermography[9], or frequency-domain signals with a single-frequency lock-in technique[10], are applied to heat the sample. Pulsed phased infrared thermography (PPT) combines simultaneously advantages of pulsed and lock-in thermography[11]. Besides a larger depth range[12], another advantage of phase angle images is that they ignore surface structures related to optical and infrared properties[10] and therefore unwanted effects from environmental reflections, emissivity variations, or non-uniform heating are suppressed. An alternative excitation method called truncated-correlation photothermal coherence tomography (TC-PCT), which also combines the advantages of time- and frequency domain methods, was inspired by radar technology and uses a chirped pulse approach in which a broadband thermal relaxation chirp is cross-correlated with a sequence of delay-swept and pulse-width-truncated references[13].

---

a) Electronic mail: peter.burgholzer@recendt.at

Heat diffusion and the propagation of an undamped acoustic wave are mathematically modeled by the diffusion equation and the wave equation respectively. Both equations are second order linear partial differential equations in space and time but describe quite different physical phenomena. While undamped acoustic waves are invariant to time reversal and therefore can be used for ultrasound image reconstruction[14], the diffusion of heat is an irreversible process, which cannot be time-reversed. As described by Salazar[15], in contrast to an acoustic wave "thermal waves" do not have a mean energy propagation.

Damped acoustic waves can be modeled in the temporal Fourier space ($\omega$ – space) by the Helmholtz equation[16] with a complex wave number $K(\omega)$, where the imaginary part is equal to the attenuation coefficient in space[17]. La Riviere et al. [18] introduced a local transformation of the acoustic wave without damping to the actual measured signal including damping. The inverse transformation of the measured damped acoustic wave to the virtual wave without damping is ill-posed and has to be regularized in order to reconstruct acoustic images from the signals. The evolution in space and time of this virtual wave signal is described by the wave equation and, therefore, all well-known reconstruction algorithms from ultrasound imaging or from photoacoustic tomography[14,19,20] can be used to reconstruct the acoustic pressure distribution in space for earlier times. In particular, the initial pressure distribution can be reconstructed for a time just after the excitation pulse. One reconstruction method which directly uses the time invariance of the virtual wave signal is time-reversal reconstruction[14,21,22], and another one which works well for planar sample surfaces is the frequency domain synthetic aperture focusing technique (F-SAFT) [23,24,25].

In this work, we introduce the concept of virtual waves for thermographic image reconstruction for the case of short excitation pulses. However, the introduced concept can be easily transferred to other temporal excitation schemes. In Section II, a local transformation similar to that for damped acoustic waves is used to calculate a virtual wave from the measured temperature signal. From this virtual wave signal, the original temperature distribution directly after the excitation pulse can be reconstructed by using any ultrasound reconstruction method. As sample surfaces are planes, F-SAFT was used. The reconstruction capability of this method was demonstrated in one, two, and three dimensions using data obtained by simulations and experiments. The reconstruction process is a two-stage process. In the first step, a virtual wave is calculated for every location of the temperature evolution on the surface. As the virtual wave is mathematically described by the wave equation, it can be reversed in time. This feature is used in a second step in order to reconstruct the temperature distribution in space inside of samples just after the excitation pulse. The first step, calculating the virtual wave, allows for all of the algorithms developed for ultrasound reconstruction to be used for thermographic reconstruction. This local transformation is presented in Section IIA. Its inverse is ill-posed and requires regularization. The influence of the regularization on the spatial resolution is given in Section IIB. Its thermodynamic background is discussed in Appendix I, where relevant prior results are also reviewed. Section III shows simulation results in two and three dimensions, which verify these resolution limits. In Section IVA the resolution limits are verified experimentally using two small spherical steel beads embedded in epoxy, heated by a short electromagnetic pulse. In Section IVB a second experimental set-up is presented where four steal beads are embedded at various depths in an epoxy cube. The infrared camera measures the temperature evolution of three orthogonal surface planes of the cube after heating by a short electromagnetic pulse. Joint reconstruction is then used to calculate a three-dimensional representation of the original heat distribution within the sample. We coin the term "thermo-tomography" for this method, where thermo-tomography refers to measurement of the temperature evolution on different surfaces and joint reconstruction to generate a three-dimensional representation of initial heat distribution.

## II. VIRTUAL WAVE SIGNALS CALCULATED FROM LOCAL TEMPERATURE SIGNALS

Temperature signals are mathematically modeled by the heat diffusion equation, whereas undamped acoustic waves are modeled by the wave equation. Linearity assures that all solutions of both equations can be represented in the temporal Fourier space ($\omega$ – space) as a superposition of "wave trains" having a certain frequency $\omega$. The "thermal wave" is a solution of the heat diffusion equation in $\omega$ – space for a certain frequency $\omega$, which is Fourier-transformed back to real space. From the linearity of the diffusion equation, it is evident that such "heat wave trains" show an interference behavior[26], as every superposition of them is again a solution of the diffusion equation. For acoustic waves, this interference behavior is widely accepted.

To calculate the virtual wave signals from local temperature signals we introduce a similar local transformation as used by La Riviere et al.[18] for damped acoustic waves (see Section I). In this work, the Helmholtz equation was not solved directly, but a relation between the signal in the presence of attenuation to an ideal signal, as would have been obtained in the absence of attenuation, was derived. This ideal pressure signal is not a real pressure wave, but only a mathematical construct, which is used for the reconstruction of the initial pressure distribution. The real measured pressure wave is the attenuated signal and in an experimental set-up one cannot only measure the ideal signal without the attenuation.

Here, such a relation between the temperature signal described by the heat diffusion equation and a virtual temperature signal is derived. The evolution in space and time of this virtual temperature wave signal is described by the wave equation and, therefore, all the reconstruction methods known from ultrasound imaging can be applied to reconstruct the temperature distribution for a time just after the excitation pulse. Like the ideal wave in an acoustic lossy medium, this temperature wave signal has no real physical realization. It is only used for mathematical reconstruction and is therefore 'virtual'.

Ammari[27] derived the same relation as La Riviere et al.[18] directly from the equations in the $\omega$ – space. In this Section, this compact derivation is presented for the heat diffusion equation.

## A. Local transformation in the temporal frequency domain and in real space

Thermal diffusion can be described by the diffusion equation[28]

$$\left(\nabla^2 - \frac{1}{\alpha}\frac{\partial}{\partial t}\right)T(\boldsymbol{r},t) = -\frac{1}{\alpha}T_0(\boldsymbol{r})\delta(t), \tag{1}$$

where $T(\boldsymbol{r},t)$ is the temperature as a function of space and time, $\nabla^2$ is the Laplacian (second derivative in space), and $\alpha$ is the thermal diffusivity which is assumed to be homogeneous in the sample. The source term on the right side of Eq. (1) ensures that the temperature just after the short excitation pulse, which is modeled by the temporal Dirac delta function $\delta(t)$, is $T_0(\boldsymbol{r})$. This description of the thermal diffusion is based on Fourier's law, which is valid for macroscopic samples, where the propagation distance is much larger than the phonon mean free path[29].

The wave equation describes the acoustic pressure $p(\boldsymbol{r},t)$ as a function of space $\boldsymbol{r}$ and time $t$ and can be written as[30,31]

$$\left(\nabla^2 - \frac{1}{c^2}\frac{\partial^2}{\partial t^2}\right)p(\boldsymbol{r},t) = -\frac{1}{c^2}\frac{\partial}{\partial t}p_0(\boldsymbol{r})\delta(t), \tag{2}$$

where $c$ is the speed of sound and $p_0(\boldsymbol{r})$ is the initial pressure distribution just after the short excitation pulse. Now a virtual wave $T_{virt}(\boldsymbol{r},t)$ is defined in a way that the same wave equation is valid with the initial temperature distribution $T_0(\boldsymbol{r})$ and an arbitrarily chosen $c$, usually set equal to one:

$$\left(\nabla^2 - \frac{1}{c^2}\frac{\partial^2}{\partial t^2}\right)T_{virt}(\boldsymbol{r},t) = -\frac{1}{c^2}\frac{\partial}{\partial t}T_0(\boldsymbol{r})\delta(t), \tag{3}$$

$\theta$ is the temperature signal in $\omega$ – space, calculated by the temporal Fourier transform :

$$\begin{aligned}\theta(\boldsymbol{r},\omega) &= \int_{-\infty}^{\infty}T(\boldsymbol{r},t)exp(-i\omega t)dt \\ T(\boldsymbol{r},t) &= \frac{1}{2\pi}\int_{-\infty}^{\infty}\theta(\boldsymbol{r},\omega)exp(i\omega t)d\omega,\end{aligned} \tag{4}$$

and similar $\theta_{virt}(\boldsymbol{r},\omega)$ is the Fourier transform of the virtual wave $T_{virt}(\boldsymbol{r},t)$. Taking the Fourier transform according Eq. (4) of Eq. (1) results in:

$$(\nabla^2 - \sigma(\omega)^2)\theta(\boldsymbol{r},\omega) = -\frac{1}{\alpha}T_0(\boldsymbol{r}) \; with \; \sigma(\omega)^2 \equiv \frac{i\omega}{\alpha} \tag{5}$$

and of Eq. (3) gives:

$$(\nabla^2 + k(\omega)^2)\theta_{virt}(\boldsymbol{r},\omega) = -\frac{i\omega}{c^2}T_0(\boldsymbol{r}) \; with \; k(\omega) \equiv \frac{\omega}{c} \tag{6}$$

which are two Helmholtz equations: Eq. (6) with a real wavenumber $k(\omega)$ and Eq. (5) with a complex wavenumber $\sigma(\omega)$. Replacing $\omega$ by $-ic\sigma(\omega)$ in Eq. (6) gives Eq. (5) by identifying

$$\theta(\boldsymbol{r},\omega) = \frac{c}{\alpha\sigma(\omega)}\theta_{virt}(\boldsymbol{r},-ic\sigma(\omega)). \tag{7}$$

Eq. (7) is the sought relation between the measured temperature signal and the virtual wave signal in $\omega$ – space. It is a local transformation, as the location $\boldsymbol{r}$ is the same for $\theta_{virt}$ and $\theta$, and it is valid in all dimensions. Transformation back from $\omega$ – space to the time domain (inverse Fourier transform) by using Eq. (4) results in:

$$T(\boldsymbol{r},t) = \frac{1}{2\pi}\int_{-\infty}^{\infty}\frac{c}{\alpha\sigma(\omega)}\theta_{virt}(\boldsymbol{r},-ic\sigma(\omega))\exp(i\omega t)d\omega,$$
$$\text{with } \theta_{virt}(\boldsymbol{r},-ic\sigma(\omega)) = \int_{-\infty}^{\infty}T_{virt}(\boldsymbol{r},t')\exp(-c\sigma(\omega)t')dt', \tag{8}$$

which can be integrated analytically in $\omega$ and can be written as:

$$T(\boldsymbol{r},t) = \int_{-\infty}^{\infty}T_{virt}(\boldsymbol{r},t')K(t,t')dt',$$
$$\text{with } K(t,t') \equiv \frac{c}{\sqrt{\pi\alpha t}}\exp\left(-\frac{c^2 t'^2}{4\alpha t}\right) \text{ for } t > 0. \tag{9}$$

Eq. (9), like Eq. (7) connects the temperature signal to the virtual wave signal at the same location $\boldsymbol{r}$, but in time domain instead of the temporal frequency domain. It can be discretized to produce a matrix equation

$$\boldsymbol{T} = \boldsymbol{K}\,\boldsymbol{T}_{virt}, \tag{10}$$

where $\boldsymbol{T}$ and $\boldsymbol{T}_{virt}$ are the vectors of the measured temperature signal and of the virtual wave signal at discrete time steps, respectively. $\boldsymbol{K}$ is the matrix at these time steps calculated from Eq. (9). Eq. (10) can be inverted only with appropriate regularization, as the matrix $\boldsymbol{K}$ shows a deficient in rank. The truncated singular value decomposition (SVD) method is used to get the reconstructed signal $\boldsymbol{T}_{rec}$ as an estimate for $\boldsymbol{T}_{virt}$ from the thermographic signal $\boldsymbol{T}$ [32]. The truncation value for the smallest singular values is $1/SNR$, with $SNR$ being the signal-to-noise ratio.

## B. Fluctuations and image degradation during heat diffusion

The inversion of matrix Eq. (10) is ill-conditioned, meaning that small fluctuations in $\boldsymbol{T}$ are highly amplified when calculating $\boldsymbol{T}_{rec}$ as an estimate for $\boldsymbol{T}_{virt}$. It needs to be re-formulated for numerical treatment by including additional assumptions, such as limiting the variation of the solution[33]. This mathematical reformulation is known as regularization, including truncated SVD or Tikhonov regularization[34]. The choice of a proper regularization parameter is critical, e.g., for the truncated SVD method this is the smallest singular value taken into account.

As reviewed in Appendix I, the best reconstruction possible in temporal frequency domain ($\omega$-space) is obtained by using frequencies up to a truncation frequency $\omega_{trunc}$. A sharp pulse in $T_{virt}$, ideally a Dirac delta pulse, will keep its shape, as the wave equation Eq. (3) has only solutions without dispersion. In $\omega$-space, $\theta_{virt}$ will also keep high frequencies after the pulse travels some distance $r \coloneqq ct'$. From the delta pulse in $T_{virt}$ the corresponding temperature signal $T$ will get broadened with increasing $r$ according to Eq. (9). In frequency domain this corresponds to $\theta(r,\omega) = \frac{1}{2\alpha\sigma(\omega)}exp(-\sigma(\omega)r) = \frac{1}{2\sqrt{i\omega\alpha}}exp(-\sqrt{i\omega/\alpha}\,r)$, as can be seen by applying the Fourier transform (4)[35]. The temperature signal in $\omega$-space is a superposition of the spectral components of "thermal waves" with angular frequency $\omega$ and wavenumber $k(\omega) := \sqrt{\omega/2\alpha}$. At a distance $r$ for a certain frequency $\omega$ the amplitude is damped by $exp(-r\sqrt{\omega/2\alpha})$. A "thermal wave" with frequency $\omega_{trunc}$ is damped at a distance $r$ by the factor $exp(-r\sqrt{\omega_{trunc}/2\alpha})$ just to the noise level, which gives

$$\omega_{trunc} = 2\alpha\left(\frac{\ln(SNR)}{r}\right)^2, \tag{11}$$

with the signal-to-noise ratio $SNR$ and ln is the natural logarithm. By applying the inverse Fourier transform Eq. (4) up to the frequency $\omega_{trunc}$ it follows that the reconstructed temperature signal $T_{rec}$ for the initial time $t = 0$ is proportional to[36]:

$$\frac{1}{2\pi}\int_{-\omega_{trunc}}^{\omega_{trunc}}\frac{1}{2\alpha\sigma(\omega)}exp(-\sigma(\omega)r)d\omega = \frac{1}{\pi}\frac{\sin(k_{trunc}r)}{r}exp(-k_{trunc}r), \tag{12}$$

$$\text{with } k_{trunc} := \sqrt{\omega_{trunc}/2\alpha} = \ln(SNR)/r.$$

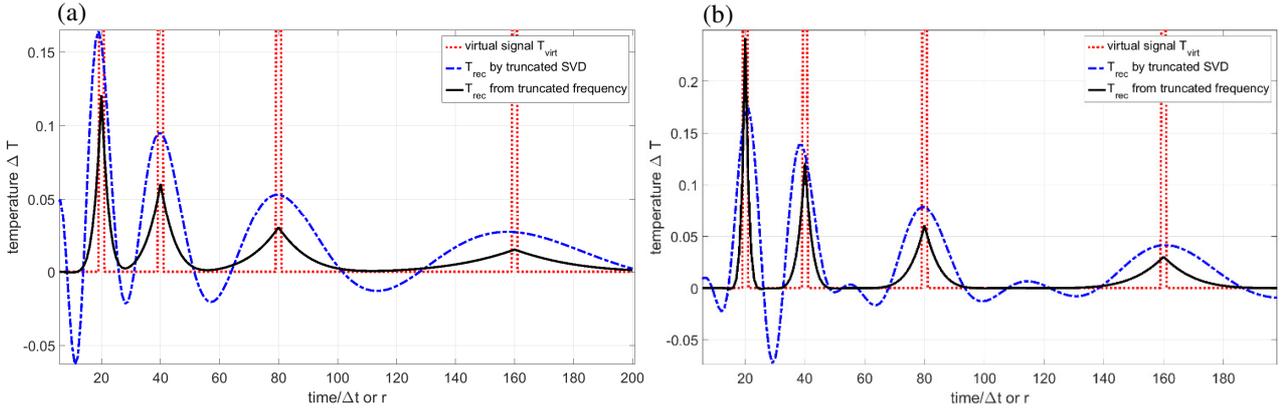

FIG. 1. Reconstructed temperature signal from a virtual signal $T_{virt}$ (dotted red line) with peaks at a depth of 20, 40, 80, and 160. The reconstructed signal $T_{rec}$ is calculated by the truncated SVD method, described in Section A (dashed blue line), and by using the truncation frequency $\omega_{trunc}$ and Eq. (12) (bold black line). The resolution $\mu$ after some time $t$ (or some travelling distance $r = ct$) is the width of $T_{rec}$ using $\omega_{trunc}$ and agrees with the "wavelength" at $\omega_{trunc}$ (zero points of the reconstructed signal in Eq. (12)). The signal-to-noise ratio $SNR$ was chosen 2000 in (a) or 2000² in (b).

In FIG. 1 this reconstructed signal $T_{rec}$ is shown for several delta peaks of the virtual signal $T_{virt}$ and compared to the reconstructed signal calculated by the truncated SVD method described in Section A for two different $SNR$ values. As the resolution limit is the size of a small (ideally point-like) reconstructed structure at a certain depth, the resolution $\mu$ after some time $t$, or some travelling distance $r = ct$, is the width of the reconstructed signal calculated by the truncated SVD method (dashed line in FIG. 1). This is in good agreement with the distance of the zero points of the reconstructed signal in Eq. (12):

$$\mu = \frac{2\pi}{\sqrt{\omega_{trunc}/2\alpha}} = 2\pi\frac{r}{\ln(SNR)} \tag{13}$$

The resolution limit is proportional to the travel distance $r$ of the signal between excitation location and detection point and inversely proportional to the natural logarithm of the signal-to-noise ratio. The resolution does not depend on the thermal diffusivity $\alpha$. For a $SNR$ value of 2000 (FIG. 1(a)) the resolution is approximately 17, 33, 66, and 132 for a travel time or distance $r$ of 20, 40, 80, and 120, respectively. For the squared $SNR$ value of $2000^2$ the resolution $\mu$ is found to be 8.5, 16.5, 33, and 66, according to Eq. (13) (FIG. 1(b)).

## III. SIMULATION RESULTS FOR 2D AND 3D HEAT DISTRIBUTIONS WITH ADIABATIC BOUNDARY CONDITIONS

The one-dimensional solution of the wave equation is a wave package of constant shape travelling with sound velocity $c$ (already published in 1747 by d'Alembert). Therefore, in one dimension (1D) the virtual wave immediately gives the reconstructed image at time $t = 0$ by projecting it back a distance $ct$. For higher dimensions, at a certain time $t$ the signals from all locations at the distance $ct$, which is a circle or sphere, interfere and summate. Reconstruction methods known from ultrasound imaging like spherical backprojection[19] (or other methods as mentioned in Section I) are used as a second step for image reconstruction from the calculated virtual wave signals. In this Section the sample is assumed to be thermally isolated, which results in adiabatic boundary conditions where the heat flux and, therefore, the normal derivative of the temperature $T(\boldsymbol{r}, t)$ at the sample boundary vanishes at all times $t$. As no heat can flow in or out of the sample, after a long time a thermal equilibrium will be established at an equilibrium temperature, which is above the ambient temperature by an amount equal to the energy of the heating pulse divided by the heat capacity and the mass of the sample.

For the numerical examples shown in this Section the Fourier transform in space was used to determine $T(\boldsymbol{r}, t)$ from the initial temperature distribution $T_0(\boldsymbol{r})$[35,36]. In this Fourier space, called "$k$-space" corresponding to $\omega$-space for the temporal frequency domain, the temperature evolution in time is a simple multiplication by $\exp(-k^2 \alpha t)$, with wave vector $\boldsymbol{k}$[37] and thermal diffusivity $\alpha$, which was assumed to be homogenous in the sample (see Eq. (1)). For adiabatic boundary conditions, the eigenfunctions are cosine functions[38] with a zero gradient at the sample boundaries. Therefore, first the discrete cosine-transform $\widehat{T}_0(\boldsymbol{k})$ from $T_0(\boldsymbol{r})$ is calculated. Then for a later time $t$:

$$\widehat{T}(\boldsymbol{k}, t) = \widehat{T}_0(\boldsymbol{k}) \exp(-k^2 \alpha t). \qquad (14)$$

The inverse cosine transform gives $T(\boldsymbol{r}, t)$, which is evaluated at the sample surface points $\boldsymbol{r}_S$ to simulate the measured surface temperature $T(\boldsymbol{r}_S, t)$. Added Gaussian noise provides the desired signal-to-noise ratio. As a numerical example for an initial temperature distribution $T_0$ three small Gaussian peaks were chosen on a two-dimensional (2D) 200 x 100 grid (FIG.

2(a)). To calculate the discrete cosine transform $\widehat{T}_0$, the time evolution $\widehat{T}$ according to Eq. (14), and the inverse discrete cosine transform to determine the surface temperature $T(\boldsymbol{r}_S, t)$ (FIG. 2(b)), a short MATLAB routine was written. More details are given in Sec. IIIA.

For every detection point $\boldsymbol{r}_S$ the truncated SVD method (Section IIA) was used to calculate the reconstructed virtual wave signal $T_{rec}(\boldsymbol{r}_S, t)$. $T_{rec}$ is the estimated virtual wave which is used in 2D and 3D to reconstruct the initial temperature distribution $T_{0,rec}(\boldsymbol{r})$. As detection surfaces for the simulations and experiments are planes F-SAFT[23,24,25], a well-known ultrasound image reconstruction method was used, but other methods such as time-reversal reconstruction also show similar results[14].

### A. Two-dimensional reconstructions from simulated data

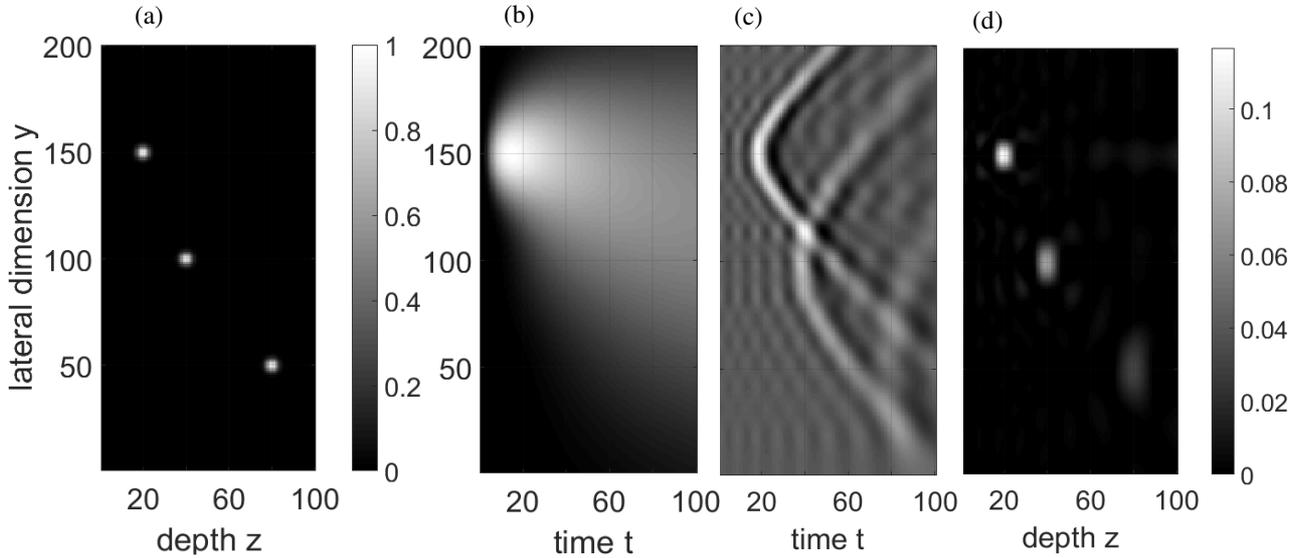

FIG. 2. 2D reconstruction example: (a) cross Section perpendicular to the surface ($z = 0$) of the initial temperature distribution $T_0(y, z)$, consisting of three small Gaussian peaks, centered at (150,20), (100,40), and at (50,80). (b) "Measured" surface temperature $T(y, t)$ at depth $z = 0$, calculated in $k$ – space using Eq. (14) from the initial temperature distribution in (a). (c) Virtual wave signal $T_{rec}(y, t)$ calculated by the truncated SVD method (Section IIA). (d) Reconstruction of the initial temperature distribution $T_{0,rec}(y, z)$ using F-SAFT.

FIG. 2 gives a 2D example for a reconstruction from simulated data, with a lateral dimension $y$ (length 200) and the depth $z$ (maximum depth 100). The initial temperature distribution $T_0(y, z)$ and its temporal evolution according to the heat diffusion equation were calculated on this 200 x 100 grid, using adiabatic boundary conditions. As described in the beginning of Section III, the adiabatic boundary conditions were implemented by calculating the 2D discrete cosine transform $\widehat{T}_0(k_y, k_z)$ from $T_0(y, z)$. For a later time $t = 1, 2, \dots 100$ $\widehat{T}(k_y, k_z, t)$ is calculated in $k$ – space by multiplication with $\exp(-k^2 \alpha t)$ according to Eq. (14). The thermal diffusivity $\alpha$ was chosen equal to 10. The inverse 2D discrete cosine transform gives the temperature distribution $T(y, z, t)$ on the 200 x 100 numerical grid. For reconstruction, only the surface temperature at $z = 0$ is taken: $T(y, t) = T(y, z = 0, t)$. Experimentally such a 2D heat diffusion could be realized e.g. by embedding thin parallel wires in a sample, which are heated by a short electrical pulse. FIG. 2 (a) shows the cross-Section into the depth of the sample of the initial temperature distribution $T_0(y, z)$ with three small Gaussian peaks with a full width at half maximum of 5 at depths $z$ of 20, 40, and 80, equal to the positions of the peaks for the 1D example in FIG. 1. The $SNR$ was chosen $2000^2$ divided by the square root of the number of measured data points (pixel number of the camera) to get a comparable result to FIG. 1 (b). FIG. 2 (b) shows the simulated "measured" surface temperature $T(y, t)$ for every lateral distance $y$ as a function of time. The corresponding virtual wave signal $T_{rec}(y, t)$ calculated by the truncated SVD method (Section IIA) is shown in FIG. 2 (c), where the parabolic shape of the virtual signal can be seen. In addition, the degradation of signals arriving later is evident, as they become more blurred at later measurement times. FIG. 2 (c) corresponds to the reconstructed image obtained if only one-dimensional thermographic reconstruction were used. This is only adequate for layered structures, but not for small inclusions. The axial resolution in the $z$ -direction after F-SAFT reconstruction in FIG. 2 (d) is the same as for the peaks at a depth of 20, 40, and 80 in FIG. 1. The width of the reconstructed peaks in the y-direction (lateral direction) is approximately twice the width in the axial direction. This effect is known from acoustic reconstruction from limited angle or limited view data[39] and can be circumvented by collecting measurement data not only from one side but also from two sides (called L-reconstruction[25]) or even more sides.

### B. Three-dimensional thermo-tomography from simulated data

FIG. 3 shows a three-dimensional example of the thermographic reconstruction of a cube with an edge length of $N = 48$ containing 4 spherical inclusions with the center points at A $(\frac{N}{4}, \frac{N}{2}, \frac{N}{2})$, B $(\frac{3N}{4}, \frac{N}{2}, \frac{N}{2})$, C $(\frac{N}{2}, \frac{N}{4}, \frac{N}{4})$, and D $(\frac{N}{2}, \frac{3N}{4}, \frac{3N}{4})$. The temperature evolution was calculated in $k$-space using Eq. (14) at three orthogonal detector planes at the cube surface with $z = N$ (top plane), $x = 0$ (right front plane), and $y = 0$ (left front plane). The three corresponding F-SAFT reconstructions are shown in FIG. 3 (b), (c), and (d), respectively. A $SNR$ of 1000 was chosen for each detector pixel - added as Gaussian noise in this simulation. As the total number of pixels for one detector plane is $N^2$, the truncation value for the truncated SVD is $1/(\sqrt{N^2}SNR) = 1/(N\,SNR)$. Therefore, the effective $SNR$ for regularization is $N\,SNR = 48000$. For all three reconstructions from the three detector planes, limited-view artifacts are visible and the inclusions far from the detector planes cannot be reconstructed at all. Better results could be obtained by using a single reconstruction incorporating data from all three detection surfaces, e.g. by using a time-reversal reconstruction algorithm[14] or reconstruction methods from Kunyansky[40]. However, as demonstrated in the past simply taking the sum of the reconstructions from the three detector planes gives similar results and gives reasonable reconstruction quality[25], which is shown in FIG. 3 (e). As mentioned in the introduction, we introduce the name "thermo-tomography" for reconstructions from thermographic signals from several surfaces, in contrast to reconstruction from a single detector plane. Thermo-tomography enables a good reconstruction quality of internal structure in three dimensions.

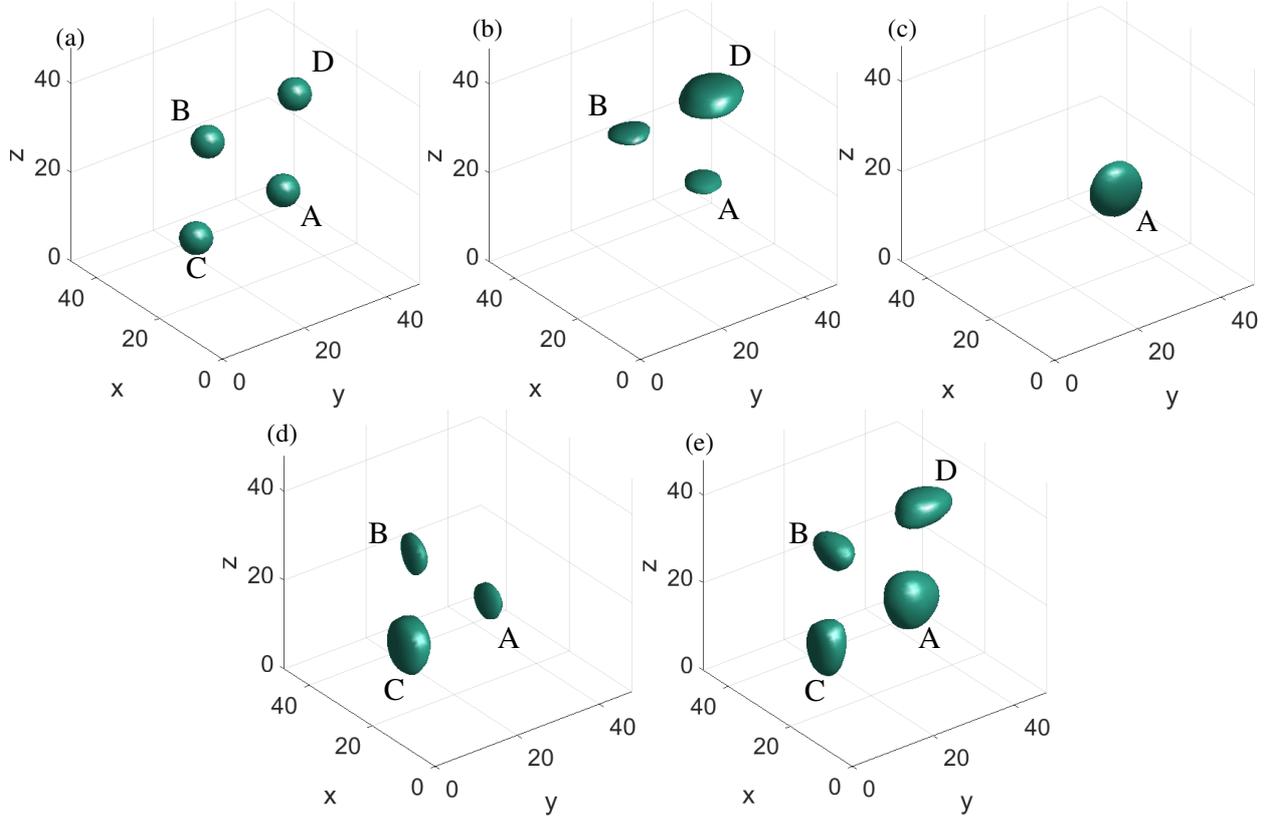

FIG. 3. Three-dimensional thermo-tomography: (a) initial temperature distribution $T_0(x,y,z)$, consisting of spherical inclusions with the center points A (12,24,24), B (36,24,24), C (24,12,12), and D (24,36,36). (b), (c), and (d) are the F-SAFT reconstructions of the initial temperature distribution for the three detector planes at the cube surface with $z = N$ (top plane), $x = 0$ (right front plane), and $y = 0$ (left front plane), respectively. (e) Reconstruction of the initial temperature distribution $T_{0,rec}(x,y,z)$ as the sum of the values from (b), (c), and (d).

## IV. THERMOGRAPHIC RECONSTRUCTION OF METALLIC INCLUSIONS IN EPOXY

### A. Experimental verification of resolution limits in 3D at various depths

Two small steel beads (diameter 2 mm, distance 7 mm) were embedded in epoxy at depths $z$ varying between 3 – 8 mm (see FIG. 4 (a)) and heated by a short electromagnetic pulse, which is known as inductive heating or induction thermography. The thermal diffusivity of a neat epoxy sample was obtained by an experimental measurement using the linear diffusivity fitting (LDF) method[41] as $\alpha = 1.3 \times 10^{-7} \frac{m^2}{s}$. In this experiment, a sample absorbs a short optical energy pulse on the front surface and the resulting temperature rise at the rear face is analyzed. The measured temperature data is fitted with the solution of the heat conduction equation for the estimation of the thermal diffusivity. The cylindrical test sample was made of an epoxy system, which polymerize at ambient temperatures when the epoxy resin (SpeciFix resin) reacts with the hardener (SpeciFix – 20 curing agent). The sample was positioned in the center of the induction coil (FIG. 4 (b), diameter 80 mm). The distance between the coil and the sample was 20 mm. The induction-heating generator had a nominal output power of 3 kW at a frequency of 220 kHz. For each depth $z$ the total measurement time $t_{meas} = 2z^2/\alpha$ was chosen. The total measurement time was determined by the Fourier number $Fo = \alpha t/z^2$. It is a dimensionless number that characterizes transient heat conduction[42]. A Fourier number of $Fo = 2$ ensures that the measurement time is long enough to reach the temperature maximum at the surface of the specimen. In 3D the temperature maximum occurs at a time $t_{max} = z^2/(6\alpha)$[43]. The pulse length was chosen to be 2% of $t_{meas}$ to approximately resemble a Dirac delta pulse in time, which was the assumption made in Section II. Basing the pulse length and, therefore the pulse energy on $z^2$ ensures that the temperature rise on the epoxy surface is approximately the same for all the samples for different depths $z$.

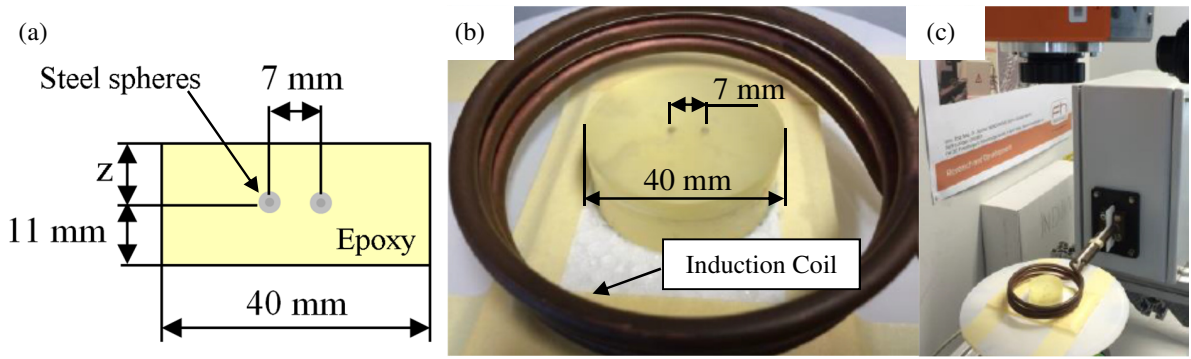

FIG. 4. Experimental set-up: (a) two small steel beads (diameter 2 mm, distance 7 mm) were embedded in an epoxy cylinder with a diameter of 40 mm at a depth z varying between 3 – 8 mm. (b) The beads were heated by a short pulse by a surrounding induction coil. (c) The surface temperature was measured by an infrared camera at a distance of 235 mm above the epoxy surface.

The infrared camera (IRcam EQUUS 81k M pro) was positioned 235 mm above the epoxy sample to keep the influence of the induction coils on the camera low. The size of the resulting pixel was (0.29 x 0.29) mm². The infrared camera uses an indium antimonite focal plane array (FPA) detector (320 x 256 pixels) which is sensitive to radiation in the 3 – 5 micron spectral range. In this spectral range, the epoxy material is opaque and the camera detects only the infrared emission of the surface. The camera has a noise equivalent differential temperature (NEDT) smaller than 20 mK. The integration time was 1 ms. Because of the short pulse time (TAB. 1) and the small beads, the temperature rise on the epoxy surface was about 300 mK (FIG. 5 (a), (b), and (c)). Therefore, the $SNR$ for one camera pixel was only about 15. For the reconstruction, the central $NxN$ pixels of the camera with $N = 70$ have been taken to avoid effects from side boundaries. This central surface area of 20 mm x 20 mm size is big enough for thermographic measurements. Outside this area, the temperature rise is below the noise level (FIG. 5 (a), (b), and (c)). The truncation value for the smallest singular values is $1/(SNR)$ (see end of Section IIA). For $NxN$ data points (camera pixels) used for the reconstruction the $SNR$ is enhanced by a factor equal to the square-root of used data points and the effective $SNR$ for regularization is $N\ SNR = 1050$. The resolution from Eq. (13) is

$$\mu = 2\pi \frac{z}{\ln(N\ SNR)}. \tag{15}$$

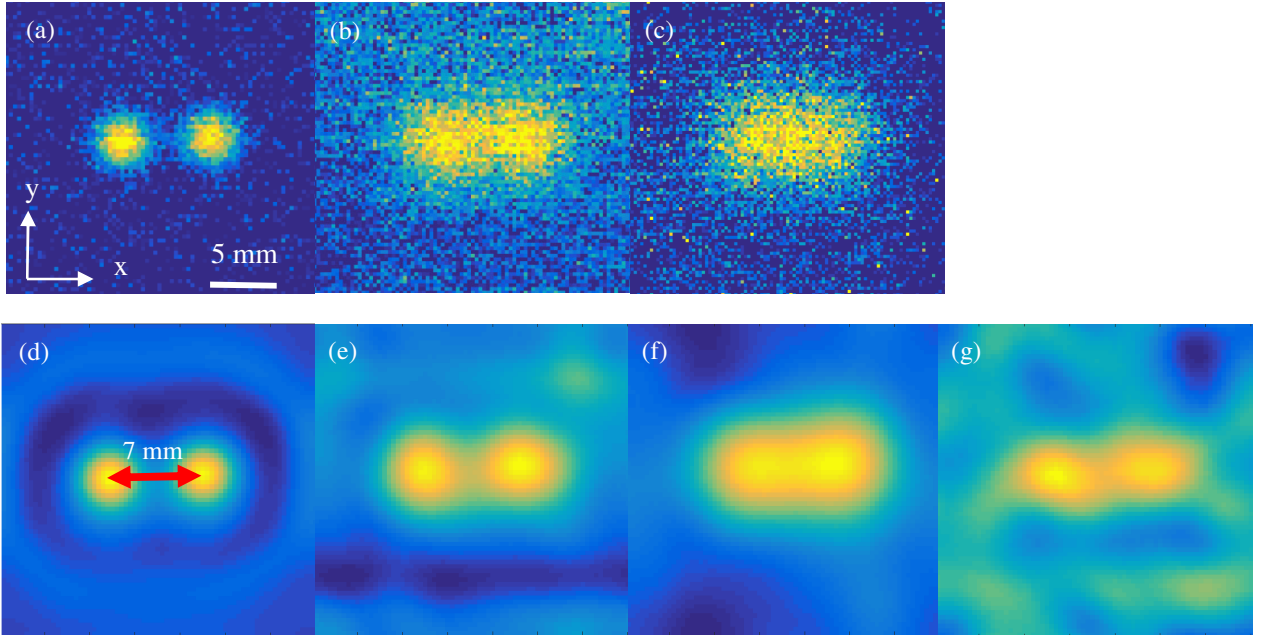

FIG. 5. Top: Temperature signal from thermographic measurement of two steel beads parallel to the epoxy surface at a depth $z$ of 3 mm (a), 6 mm (b), and 8 mm (c) with highest signal contrast at times of 7s, 26 s, and 40 s, respectively. Bottom: Thermographic image reconstruction of the two steel beads parallel to the epoxy surface at a depth $z$ of 3 mm (d), 6 mm (e), and 8 mm (f). In (f) the resolution of 7.2 mm is higher than the distance of the beads and they cannot be resolved. (g) Artifacts come up when more singular values are taken to calculate the pseudo-inverse matrix $\mathbf{K}^+$ than allowed by the $SNR$ (see text).

FIG. 5 (d), (e), and (f) show the reconstructions at depths of the steel beads $z$ of 3 mm, 6 mm, and 8 mm. The resolution according to Eq. (15) is 2.7 mm, 5.4 mm, and 7.2 mm, respectively. As the distance of the two beads is 7 mm they cannot been resolved in (f). Artifacts come up when more singular values are taken to calculate the pseudo-inverse than allowed by the $SNR$. Using 15 singular values would correspond to an $SNR$ of 7000 instead of the 12 singular values corresponding to the actual effective $SNR$ of 1050. This gives a better resolution according to Eq. (15) of 5.7 mm and the two beads can still be resolved as shown in FIG. 5 (g). However, additional artifacts come up because of the influence of the fluctuations. The corresponding measurement parameters are listed in Tab. 1.

TAB. 1. Measurement parameters of the thermographic experiment for the reconstruction of the steel beads at a depth of 3 mm, 6 mm and 8 mm.

| Depth (mm) | Measurement time (s) | Pulse duration (s) | Frame rate (Hz) |
|---|---|---|---|
| 3 | 52 | 1 | 25 |
| 6 | 208 | 4 | 7 |
| 8 | 371 | 7 | 4 |

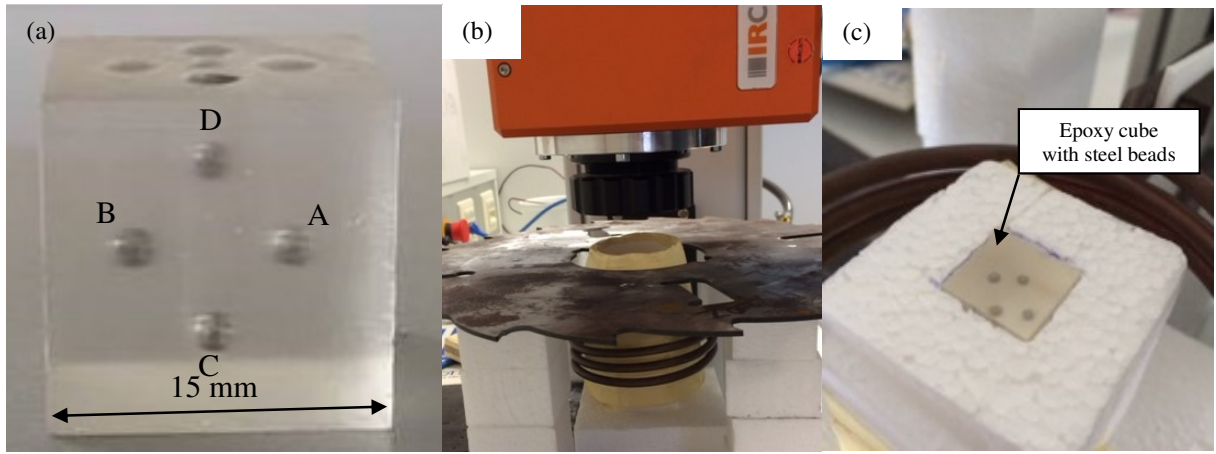

FIG. 6. Experimental set-up for 3D thermo-tomography: (a) 4 spherical steel beads (diameter 2 mm) were embedded in an epoxy cube with an edge length of 15 mm as in the simulations shown in FIG. 3 (a). The front plane, the top plane, and the left side plane were measured and used for tomographic reconstruction. (b) Measurement set-up similar to FIG. 4, only the distance of the infrared camera was reduced to 95 mm to achieve 140 pixels for the edge length of the cube (9.33 pixels per mm). The camera was therefore protected by a metal sheet from the influence of the induction coils. (c) The bottom and the sidewalls of the cube were thermally isolated using polystyrene foam to ensure adiabatic boundary conditions.

## B. Thermo-tomography of small metallic inclusions in an epoxy cube

The same experimental set-up as in FIG. 4 was used to measure the thermographic signals from four spherical steel beads with a diameter of 2 mm, which were embedded in an epoxy cube with an edge length of 15 mm (see FIG. 6 (a)). The geometry corresponds to the simulation shown in FIG. 3. The epoxy material, the steel beads, the inductive heating, and the position of the sample between the coils were the same as in Section A. Only the distance between the front lens of the infrared camera and the sample surface was reduced to 95 mm to achieve 140 pixels for the cube edge length of 15 mm (FIG. 6 (b)). The bottom and the sidewalls of the cube were thermally isolated using polystyrene foam to ensure adiabatic boundary conditions (FIG. 6 (c)). The temperature evolutions on the front plane, the top plane, and the left side plane were measured and used for tomographic reconstruction. As the steal beads had depths $z$ of approximately 4.3 mm, 7.5 mm, and 10.7 mm from the measurement planes, it was not possible to adjust the pulse times for every individual depth as in Section A. The pulse time 7.5 s was chosen to obtain thermographic signals from the deeper beads. For the bead at a depth of 4.3 mm, the maximum of the thermographic signal is at a time $t_{max}$ of 23.7 s. Therefore, for this bead the pulse time is nearly one third of $t_{max}$, and the assumption of the Dirac delta pulse excitation is only approximately valid any more. Using this Dirac delta pulse excitation assumption, the reconstruction in FIG. 7 still works well, but the position is slightly shifted to higher depths.

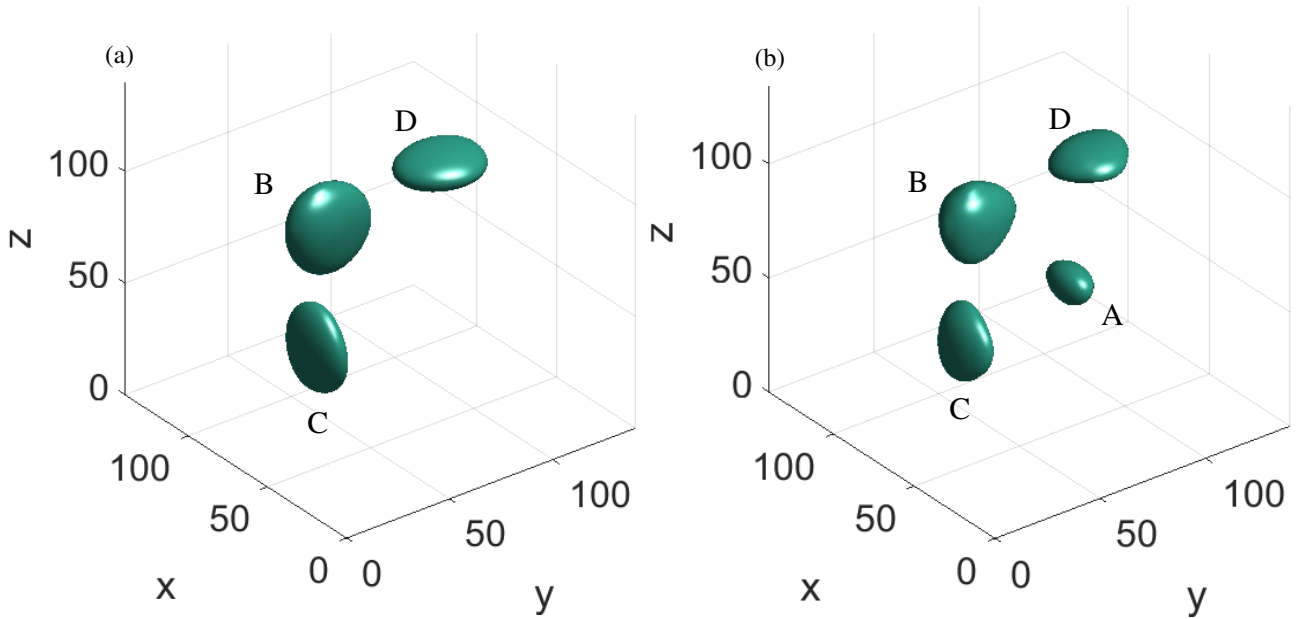

FIG. 7. Reconstruction results for 3D thermo-tomography: (a) sum of the reconstructions from top plane, left front, and left back plane measurements. Due to the poor SNR (see text) only the steal beads near to the measurement planes at a distance of 4.3 mm (approximately 40 pixels) could be reconstructed. Due to limited view artifacts (see also FIG. 2) the lateral resolution is worse than the axial resolution (ellipsoidal shape) and this effect is not reduced by adding reconstructions from other measurement planes, because the resolution is too bad for the other measurements to have a significant influence. (b) Filtering the measurement data (see text) before reconstruction allows the reconstruction of the fourth steal bead and the artifacts due to limited view are better compensated.

The sampling rate of the camera was chosen at 10 Hz and the distance to the cube surface was reduced to 95 mm to obtain more pixels for the reconstruction ($N = 140$, FIG. 6 (b)). The *SNR* for the thermographic signals was about 90, which gave an effective *SNR* for the regularization with $N\ SNR = 12\ 600$. For shorter pulse times the *SNR* was worse. This *SNR* is sufficient to achieve a good reconstruction of the steal beads near the measurement planes at a depth of 4.3 mm (approximately 40 pixels), shown in FIG. 7 (a). To avoid limiting-view artifacts (ellipsoidal shape of the reconstructed spheres) in thermo-tomography a higher *SNR* is necessary. This is different from the limiting-view artifacts in pure acoustic imaging, where the acoustic signals are not attenuated and measurement signals from other projection directions always reduce the limiting view artifacts. The knowledge, that the first bead is at a depth of 4.3 mm, gives a truncation frequency of 0.045 Hz using Eq. (11). This truncation frequency was used as the cutoff frequency of a $6^{th}$-order low-pass Butterworth filter, which significantly increased the *SNR* of the filtered measurement signals to 1340, resulting in an effective *SNR* for the regularization with $N\ SNR = 187\ 600$. Filtering the measurement data does not affect the reconstructions at a higher depth than 4.3 mm, because the truncation frequency according to Eq. (11) is less than 0.045 Hz. The reconstruction from the filtered data is shown in FIG. 7 (b). Now a contour level could be found where the bead A is also reconstructed and the artifacts due to limited view are better compensated. For both figures in FIG. 7 the contour level is the maximum level divided by the Eulers number e.

## V. DISCUSSION AND OUTLOOK

In the presented work signal reconstruction for short excitation pulses was demonstrated. In future work the concept of virtual waves will be transferred to other excitation schemes, including longer pulses, frequency-domain excitation with a single-frequency, or chirped pulses. For simplicity, in this work, internal metallic structures were directly heated by using an induction coil. In future this method will be extended to heating the surface of opaque samples directly by the excitation pulse. The generated heat then diffuses into the sample and this diffusion will be influenced by the internal structures. Virtual waves can also be used for image reconstruction in this case. However, it is expected that the resolution limit will be twice the limit given in Eq. (13), as the heat has to diffuse from the surface to the internal structures and back again. Direct and indirect heating can be seen as the thermographic equivalent of photoacoustic imaging and laser ultrasonics. In photoacoustic imaging, ultrasonic waves are generated in internal structures by means of laser irradiation, whereas in laser ultrasonics the ultrasonic waves generated at a sample surface are scattered by the internal structures[14,44]. For both cases, F-SAFT reconstruction can be used. In laser ultrasonics, the propagation distances have to be doubled compared to photoacoustics, or alternatively, half of the sound velocity is taken[24]. Also, boundary conditions for the sample surface which are more realistic than the adiabatic boundary condition should be investigated in future. First results for a third kind boundary condition[43] are given in Burgholzer et.al.[34] Another assumption made in this publication, which can be changed in future work is the homogenous thermal diffusivity. For small and well-separated structures of different materials, like the small steel beads used in the presented work, the thermal surface signal is only slightly changed in comparison to a strictly homogeneous material. In the case of more inhomogeneous areas within the sample, a virtual wave concept with inhomogeneous sound velocity seems feasible to reconstruct the initial temperature distribution.

Beyond thermography, this work might lead to a better understanding of ill-posed inverse problems and the physical background of regularization. Thermodynamic fluctuations are usually only relevant for macroscopic samples, but for ill-posed inverse problems, they become highly amplified. This is the reason why information is lost in the direct forward problem and cannot be retrieved in the inverse problem. We have shown before that the lost information, quantitatively described by the Kullback-Leibler divergence (see Appendix I), is equal to the entropy production[36]. As known from thermodynamics, the direct forward process always shows an entropy increase and, therefore, the inverse problem has to be ill-posed. The lost information is just the produced entropy. Only for ideal, reversible processes is no entropy produced and, therefore, the inverse problem is well-posed and can be solved without regularization.

In the proposed method, all the irreversibility of the heat diffusion is contained in the transformation from $T$ to $T_{virt}$. The information loss during this transformation, which correlates with a reduced spatial resolution of the reconstructed images, is equal to the entropy production caused by the diffusion. This information loss causes a fundamental resolution limit for the reconstructed images that cannot be compensated by any reconstruction algorithm. As reviewed in Appendix I and shown in Section IIB for the case of a short pulse, ideally a Dirac delta pulse, the best reconstruction results can be obtained in the temporal frequency domain by using frequencies up to a truncation frequency $\omega_{trunc}$. At this frequency the "thermal wave" is damped just to the noise level.

## VI. CONCLUSION

For imaging sub-surface structures by means of pulsed thermography, the temperature distribution inside the sample has to be reconstructed from the temperature evolution measured on the sample surface. Instead of directly solving this ill-posed inverse problem, a two-stage algorithm was presented. First, a virtual signal is calculated from the measured temperature evolution for every detection point. This virtual signal is a solution of the wave equation with the same initial conditions as the heat diffusion equation. In contrast to measured temperature evolution, the virtual signal can be reversed in time and all well-established reconstruction methods known from ultrasound imaging can be used for thermographic reconstruction in a second step. The main advantage of the proposed two-stage algorithm over directly solving the inverse heat diffusion problem is that the local transformation introduced is the same for all sample geometries and dimensions. In the second step, standard procedures from ultrasonic imaging can be used.

The feasibility of this approach was demonstrated through simulations and experiments. In the experiments, steel beads embedded in an epoxy matrix were heated by a short electromagnetic pulse generated by a surrounding induction coil. The temperature evolution on the surface was then measured using an infrared camera in the spectral region between 3 and 5 μm. The obtained resolution limit turned out to be proportional to the depth of the structures and to be inversely proportional to the natural logarithm of the signal-to-noise ratio. In two and three dimensions, the lateral resolution can be degraded due to limited view artefacts if the temperature evolution is measured on one surface only. By acquiring temperature data on more than one surface, the resolution and quality of thermographic reconstructions can be enhanced. This effect was demonstrated in numerical simulations and by experiment. For the latter, the temperature evolution caused by the inductive heating of four steel beads was measured on three orthogonal surfaces of an epoxy cube.

Thermographic imaging using an infrared camera for signal detection has a big advantage compared to ultrasonic imaging using acoustic transducers: it does not need any coupling media and parallel detection of many camera pixels is possible. The main disadvantage of pulsed thermography is the degrading spatial resolution with increasing depth. This is the reason why often only one-dimensional thermographic reconstruction is performed, which can be applied for layered structures, where the lateral extension is big compared to their depth. For other structures, using a one-dimensional reconstruction gives additional artifacts, as the heat propagates not only perpendicular to the surface but also in the lateral direction. Solving directly the inverse three-dimensional heat diffusion equation is rather ambitious, even for simple geometries of embedded structures, e.g. rectangular vertical cracks[7]. Three-dimensional reconstruction methods, which are well known from ultrasonic imaging, can be used easily for any sample geometry by applying the proposed virtual wave concept.

**ACKNOWLEDGMENTS**

This work has been supported by the Christian Doppler Research Association (Christian Doppler Laboratory for Photoacoustic Imaging and Laser Ultrasonics), and by the project "multimodal and in-situ characterization of inhomogeneous materials" (MiCi) by the federal government of Upper Austria and the European Regional Development Fund (EFRE) in the framework of the EU-program IWB2020. This work was also financially supported by the TAKE OFF program of the Bundesministerium fuer Verkehr, Innovation und Technologie (BMVIT) and by the strategic economic- research program "Innovative Upper Austria 2020" of the province of Upper Austria. We thank Heinz Gresslehner, Thomas Berer, and Todd Murray for inspiring discussions and useful comments.

**APPENDIX I**

In this Appendix the main results from our work in the past on the connection of entropy production, information loss, and time reversal of heat diffusion are shortly reviewed[17,34,35,36], which are used in Section II B. In the past, rank deficient and discrete ill-posed problems like the inversion of Eq. (10) have been examined in many mathematical publications. E.g., Hansen[33] gives a good introduction and an overview about numerical aspects of linear inversion. If one is interested only in the application of virtual wave signals for thermographic reconstruction, one of these regularization methods can be used. But the choice of the optimal regularization method and of the regularization parameter is still a matter of discussion. Therefore, we have investigated in the past why the inversion of heat diffusion is ill-posed and give a short review of the results in this Appendix. The implications on the spatial resolution for thermographic reconstruction are discussed.

Temperature in thermodynamics describes the mean kinetic energy of the molecules and can be modeled in statistical physics as a random variable showing fluctuations about its mean value[45]. In macroscopic systems, these fluctuations are extremely small, but they are highly amplified in ill-posed inverse problems like heat diffusion and, therefore, thermodynamic fluctuations get significant. Despite the time invariance of the dynamics of the individual molecules the time evolution of temperature shows a direction of time, which is described by increasing entropy in the second law of thermodynamics. To take the influence of fluctuations and entropy production on thermographic reconstruction into account we have modeled heat diffusion as a stochastic process (Ornstein – Uhlenbeck process as a Gauss – Markov process) and found that the information loss is equal to the production of entropy during heat diffusion[35]. Fluctuations and entropy production are closely connected, and we showed the equality of entropy production and information loss is equivalent to a fluctuation-dissipation theorem, which represents a generalization of the famous Johnson[46] – Nyquist[47] formula in the theory of electric noise.

Of course modeling heat diffusion by a simple Gauss-Markov process is a rough idealization. Nevertheless, for macroscopic systems we could show the equality of entropy production and information loss from recent results of non-equilibrium statistical physics, where we only used the heat diffusion equation modeling the mean value evolution of the temperature without the necessity to describe the fluctuations[36]. The influence of fluctuations is implicitly taken into account by calculating the entropy production from the mean value equation as a simple linear differential equation and no sophisticated stochastic process has to be used. As the mean entropy production during the process, which is always positive, is equal to the information loss, the reconstruction of an earlier state from a later state of the process has to be an ill-posed problem. Any reconstruction algorithm cannot gain the lost information. Why does this information loss result in a degradation of the reconstructed image? High-resolution images contain more information than those with low spatial resolution, which is evident, e.g., when they are stored on a computer - they need more memory capacity. During heat diffusion a sharp pulse, ideally a Dirac delta pulse, which provides high resolution and contains all frequency components in temporal frequency domain and therefore much information, will be blurred and broadened according to Eq. (9). As the resolution limit is the size of a small (ideally point-like) reconstructed structure at a certain depth, this pulse broadening limits spatial resolution. *This broadening can be compensated by inversion only up to a certain extent limited by the information loss. The information lost due to fluctuations, which is for macroscopic samples equal to the entropy production, cannot be compensated by any reconstruction algorithm, and causes a principle resolution limit for the reconstructed images.*

This qualitative description of the degradation of the reconstructed image due to heat diffusion can be formulated quantitatively by using the relative entropy, called also Kullback – Leibler (KL) divergence[48], to measure the information content of a thermographic signal. The KL divergence $D(f||g)$ is used in information theory for testing the hypothesis that the two distributions with density $f$ and $g$ are different, and is defined as

$$D(f||g) := \int \ln\left(\frac{f(x)}{g(x)}\right) f(x) dx, \tag{A1}$$

where ln is the natural logarithm. The Chernoff – Stein Lemma states that if $n$ data generated from a distribution density $g$ are given, the probability of guessing incorrectly that the distribution density for describing the data is $f$ is bounded by the type II error $\epsilon := \exp(-nD(f||g))$, for $n$ large[48]. In that sense $D(f||g)$ can describe a "distance" between the distribution densities $f$ and $g$.

In thermal equilibrium with distribution density $p_{eq}$ in the state space the whole sample is on the same temperature. The entropy has its maximum and all the information for a thermographic reconstruction has been lost. At what time $t_{trunc}$ is the equilibrium reached? At a certain time $t$ the state with distribution density $p_t$ contains the information $k_B D(p_t||p_{eq})$, where $k_B$ is the Boltzmann constant, and for macroscopic system this is equal to the entropy production between the state at time $t$ and the equilibrium state[36]. From Chernoff – Stein Lemma it is known that if $D(p_t||p_{eq})$ gets smaller than a certain level $\ln\left(\frac{1}{\epsilon}\right)/n$ the state cannot be distinguished from equilibrium. Therefore at any time later than this truncation time $t_{trunc}$ thermal equilibrium is reached. $t_{trunc}$ depends on the type II error $\epsilon$.

In temporal frequency domain ($\omega$-space) a truncation frequency $\omega_{trunc}$ is used instead of the truncation time. The information content $k_B D(p_\omega(r)||p_{eq})$ of the diffusion signal with frequency $\omega$ at distance $r$ is equal to the entropy production[36,45]

$$\Delta S_\omega(r) = \frac{1}{2} k_B SNR^2 exp(-2r\sqrt{\omega/2\alpha}), \tag{A2}$$

with the signal-to-noise ratio $SNR$. The truncation frequency $\omega_{trunc}$ is determined according to the Chernoff – Stein Lemma when $\Delta S_\omega(r)$ gets smaller than $k_B \ln\left(\frac{1}{\epsilon}\right)/n$:

$$\omega_{trunc} = 2\alpha \left(\frac{\ln(SNR/C)}{r}\right)^2, \tag{A3}$$

where $C := \sqrt{2\ln\left(\frac{1}{\epsilon}\right)/n}$ is a constant value containing the error level $\epsilon$ and the number of measurements $n$, which scales the $SNR$. Choosing this constant value $C$ equal to one means that the amplitude for this "thermal wave" with frequency $\omega_{trunc}$ is damped at a distance $r$ by the exponential factor $exp(-r\sqrt{\omega_{trunc}/2\alpha})$ just to the noise level (FIG. A1) and leads to Eq. (11) in Section II B.

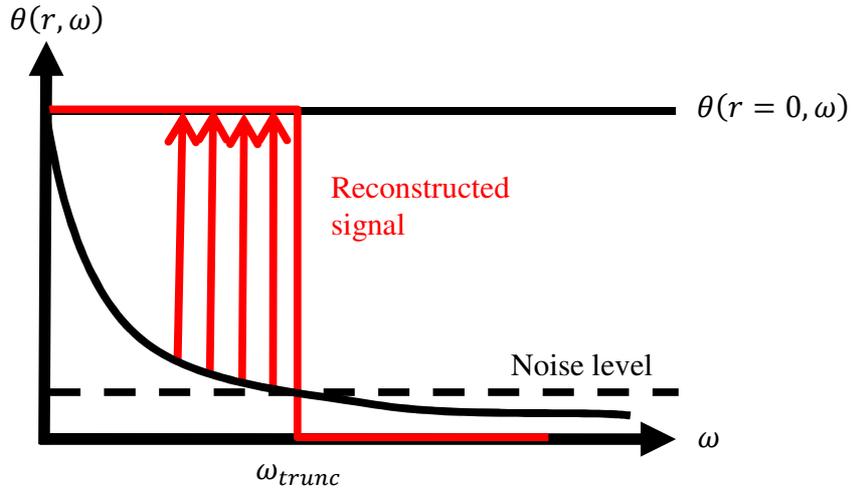

FIG. A1. Thermographic signal in $\omega$ – space: $\theta(r,\omega)$ is damped by $\exp(-r\sqrt{\omega/2\alpha})$ at a distance $r$. The amplitude with frequency $\omega_{trunc}$ is damped just to the noise level. For the reconstruction only the frequencies less than $\omega_{trunc}$ are taken.